\documentstyle[12pt,fleqn,elsevier,epsfig]{article}
\renewcommand{\baselinestretch}{1.2}

\begin{document}

\setcounter{footnote}{0}
\renewcommand{\thefootnote}{\alph{footnote}}

\vspace{2cm}
\begin{center}

{ \Large \bf Imaging Hybrid Photon Detectors with
Minimized Dead Area
and
Protection Against Positive Ion Feedback\footnote{
Submitted to Nucl. Inst. Meth. A
} \\[0.5cm]
}

\noindent
Daniel Ferenc

\noindent
Div. PPE,\\CERN,\\ 1211-Geneva,\\Switzerland\\
E-mail: Daniel.Ferenc@CERN.ch 
\end{center}

\begin{center}
Abstract
\end{center}

{
Imaging Hybrid Photon Detectors (HPD) 
have been developed for integration in 
large area Cherenkov detectors for high energy physics and astrophysics.
The presented designs -- developed particularly for
the experiments
MAGIC, LHCb and AQUA-RICH --
comprise very good imaging properties,
protection against positive ion feedback
and(or)
minimum
dead area.
The underlying innovations are discussed in some detail.
}

\vspace{1.5cm}
\renewcommand{\baselinestretch}{1.2}
\small\normalsize
\setcounter{footnote}{0}
\renewcommand{\thefootnote}{\arabic{footnote}}

\section{Introduction}
\noindent
With the onset of new technologies, 
Hybrid Photon Detectors (HPDs) became one of the
most favorable options for the detection of Cherenkov photons
in large area Ring Imaging CHerenkov (RICH)
detectors.
In contrast to ordinary 
Photo Multiplier Tubes (PMT), where
a photoelectron initiates a multi-step amplification
process in a dynode system,
in HPDs~\cite{Salvo} 
a
photoelectron 
undergoes a single-step acceleration from the photocathode
towards the silicon sensor.
Due to its relatively high energy
(5-20 keV) 
a photoelectron releases a large number of electron-hole
pairs
upon its impact 
in the semiconductor sensor.
The number of secondary electrons released in a silicon 
sensor is N$_e\simeq $(U-U$_0)/3.6$V,
where U$_0$ is the voltage (in Volts) to accelerate an electron
to an energy sufficient to penetrate the inert layer
above the sensitive semiconductor pn structure.

This publication reports on the solution of
two important problems -- the minimization of dead area
and the protection against positive ion feedback -- as implemented in concrete HPD designs.
These solutions are, however, very general and may be 
implemented (with appropriate scaling) 
in a variety of different designs.

First, we will demonstrate how to minimize the dead area of 
a large multi-channel imaging HPD in order to enable 
efficient integration of individual HPD tubes into a large area 
imaging Cherenkov detector. 
The method we apply provides 
undistorted image projection from the entire photocathode
to the surface of the pixelized semiconductor sensor. The presented 
HPD design has been 
developed~\cite{erice1,erice2} for the LHCb experiment.  Excellent imaging 
properties provided in this way 
are equally welcome in
stand-alone devices, like in 
the 20-inch HPD tube designed for the 
AQUA-RICH neutrino experiment at Gran Sasso~\cite{aquarich} --
the second design presented in this publication.

Second, 
in order to solve the positive ion feedback problem,
i.e. to prevent positive ions 
(released in impacts of photoelectrons from the surface of the semiconductor
sensor) from hitting the photocathode,
we followed
the same idea originally developed for ion feedback protection
in small 
single-channel HPDs~\cite{NIM}.
In particular, we shall demonstrate that the insertion of
a permanent potential barrier for positive ions in front of the anode 
may solve the ion feedback problem also in large multi-channel
imaging HPDs, but only if the so called cross-focusing
electron optics is applied. The ion feedback protection
has been developed for the cosmic gamma ray
Atmospheric Cherenkov Telescopes (ACT), in particular for the 
MAGIC project~\cite{MAGIC}.

For all the electron optics simulations presented,
SIMION 3D software~\cite{SIMION} has been used.
In all the figures presented, only elements 
important for electron optics (focusing electrodes)
are shown, while the supporting structures are
left out. It is assumed that the medium in the tubes is vacuum.

\section{Minimizing the dead area}
One of the most important problems in the integration of HPD tubes into
a matrix of a large-surface area RICH detector is the low overall
photo-sensitive surface coverage, caused by a typically high 
HPD dead area. HPDs have usually been designed as stand-alone
devices, and very little care has been taken of the relationship
between the physical and the sensitive surface areas.
In those applications where a pixel size
of 1-2 cm is sufficient, one can use a matrix
consisting of single-pixel HPDs.
The large dead area between the sensitive surfaces 
of the individual detectors
may be taken care of
by 
focusing the light to the sensitive photocathode areas by 
Winston cones or lenses~\cite{Winston}.
For applications which require yet a smaller pixel
size (e.g. 1 mm), this method is impossible due to the
limited possible
size of individual tubes, and therefore one should use large
diameter HPDs with internal electron optics imaging and multi-channel
silicon sensors.

The problem in the integration of these large area imaging HPDs into
hexagonal matrices is that 
it is essentially impossible to use Winston
cones to redirect 
light from the dead area into the sensitive detector area,
because that would destroy the image pattern. 
Demagnification lenses are still possible, but they would
have to be very large.

\begin{figure}[*]
\epsfig{file=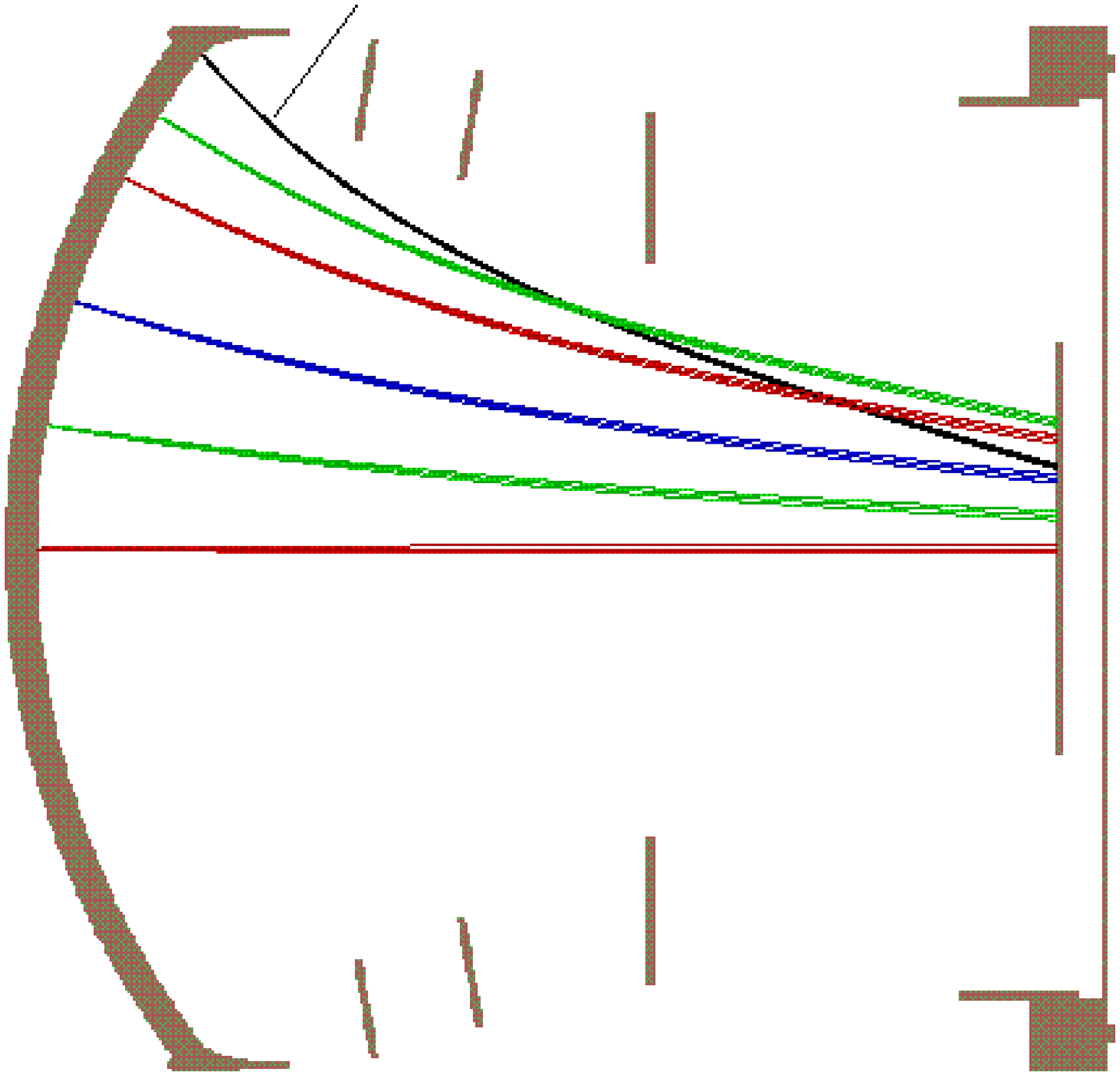,bbllx=50,bblly=150,bburx=792,bbury=620,width=22cm}
\caption{
\noindent
Proximity-focusing 5-inch diameter HPD.
Photoelectrons are
focused onto
the silicon-pad
sensor on the right side.
Photoelectrons emerging from the periphery 
of the photocathode are incorrectly focused.
Electrodes are kept at the following potentials, from left to
right, respectively: -20 kV, -15 kV, -11 kV, -4.7 kV and 0 V.
Photoelectrons are simulated with an initial energy of 0.25 eV
and an emission angle of
+45$^{\circ}$, -45$^{\circ}$ and 0$^{\circ}$ relative to the
normal.
}
\label{FPP01}
\end{figure}

With the first 
HPD design to be presented, see Fig.~\ref{FPP01},
the idea was to extend the sensitive
photocathode surface almost to the physical edge
of the device.
This 5-inch HPD was designed particularly for the
hexagonally close packed 
Ring Imaging Cherenkov detectors 
of the LHCb experiment~\cite{erice1,erice2}.
Its size and the demagnification factor were determined 
from the 
size of the 2048-channel 
silicon-pad sensor (5-cm diameter) used for the 
detection of the photoelectrons and the
required position
resolution for Cherenkov photons.
The radius of curvature of the spherically shaped entrance window
was originally taken such that the center of curvature 
was placed close to the surface of
the silicon sensor. This provides a very good mechanical
stability, needed because of the thin window connection with the
body of the tube. Further optimizations of the window curvature
for the electron optics quality
are still possible, but it turns out, as we shall demonstrate
in some detail, that a very good result may be achieved already
with the original curvature. However, the optimization of the tube
length will be considered below.

\begin{figure}[*]
\epsfig{file=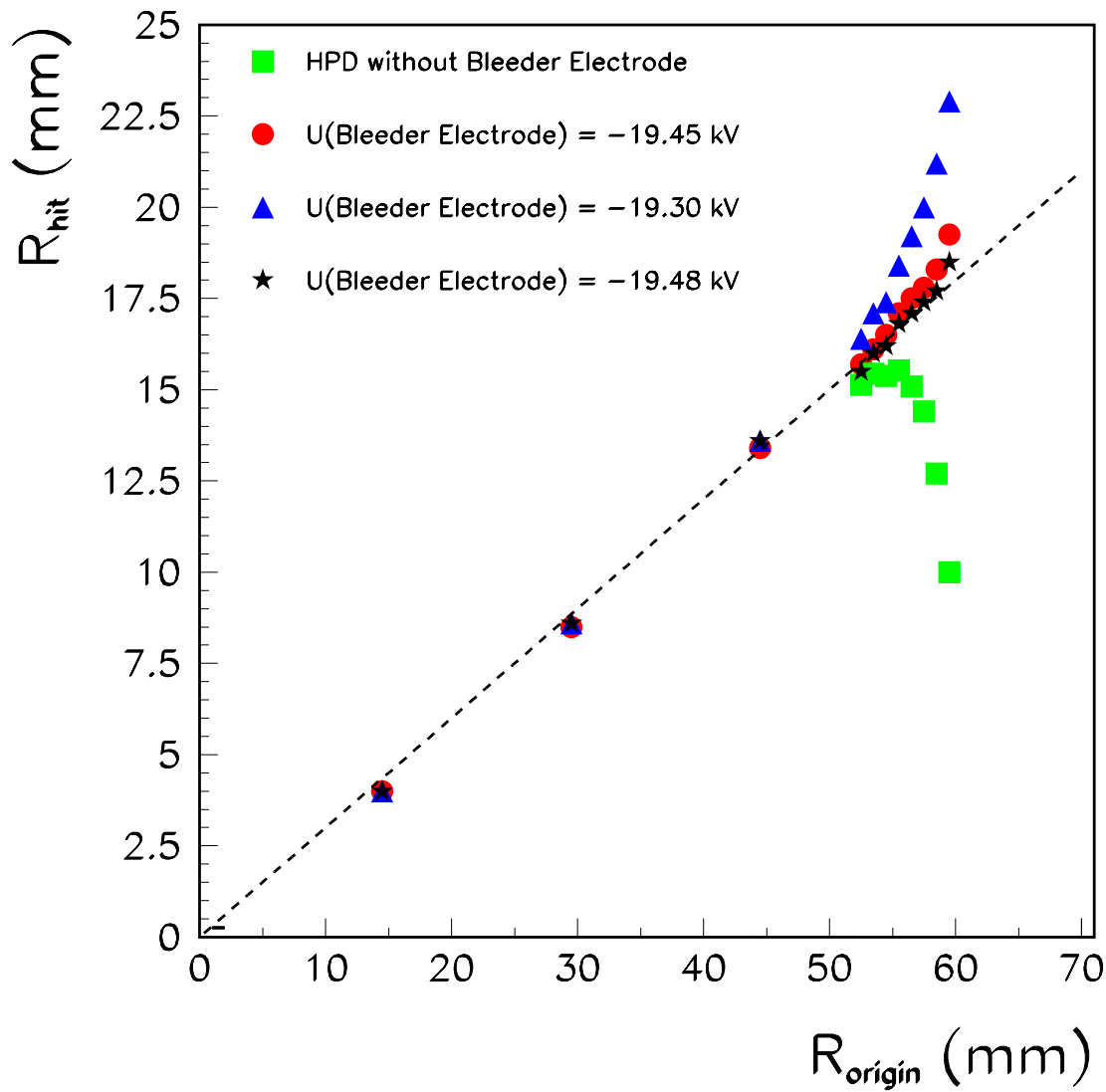,bbllx=50,bblly=200,bburx=792,bbury=480,width=22cm}
\caption{
\noindent
Mapping of the image from the photocathode to the 
silicon pad sensor. In the 
simulation electrons originate from the photocathode at a radius R$_{origin}$ 
from the HPD axis,
and hit the silicon sensor at R$_{hit}$.
Two different HPD designs are presented: without and 
with a new ``bleeder electrode". The original HPD without
the bleeder electrode (squares) fails in imaging.
}
\label{FPP011}
\end{figure}

Electron focusing is demonstrated in
Fig.~\ref{FPP01} -- photons (not shown in the figure) enter HPD 
from the left side, pass through the entrance window, and release
photoelectrons from the photocathode on the interior of the window.
Photoelectrons then undergo
acceleration and electrostatic focalization
towards the silicon-pad
sensor, shown on the right side. Some example
trajectories are presented in Fig.~\ref{FPP01}.
A relatively smooth mapping
between the image on the photocathode and the 
projected image on the silicon sensor 
has been achieved
close to the
axis of the tube but,
evidently, mapping completely fails
in the region close to the periphery of the tube where
strong image folding is
present (the square symbols in Fig.~\ref{FPP011}).
Although the physical shape of this HPD has been
successfully optimized for close packing by maximal extension of the
photocathode 
towards the periphery,
the device demonstrates
unacceptably large {\it functionally} dead area.

\begin{figure}[*]
\epsfig{file=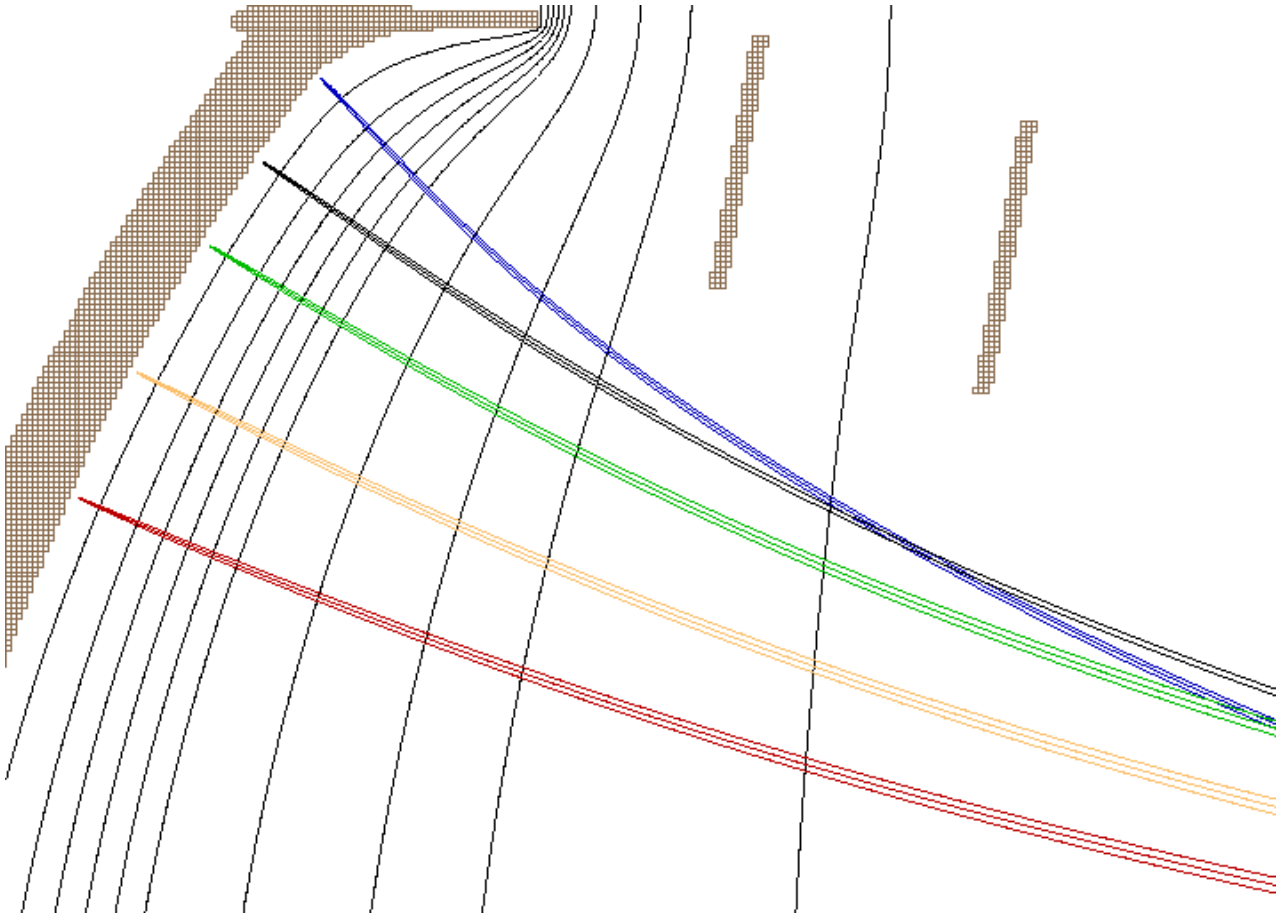,bbllx=50,bblly=200,bburx=792,bbury=480,width=24cm}
\caption{
\noindent
Incorrect electron focusing close to the 
periphery of the photocathode is due to the
bunching of equipotential surfaces near the edge.
}
\label{FPP02}
\end{figure}

\begin{figure}[*]
\epsfig{file=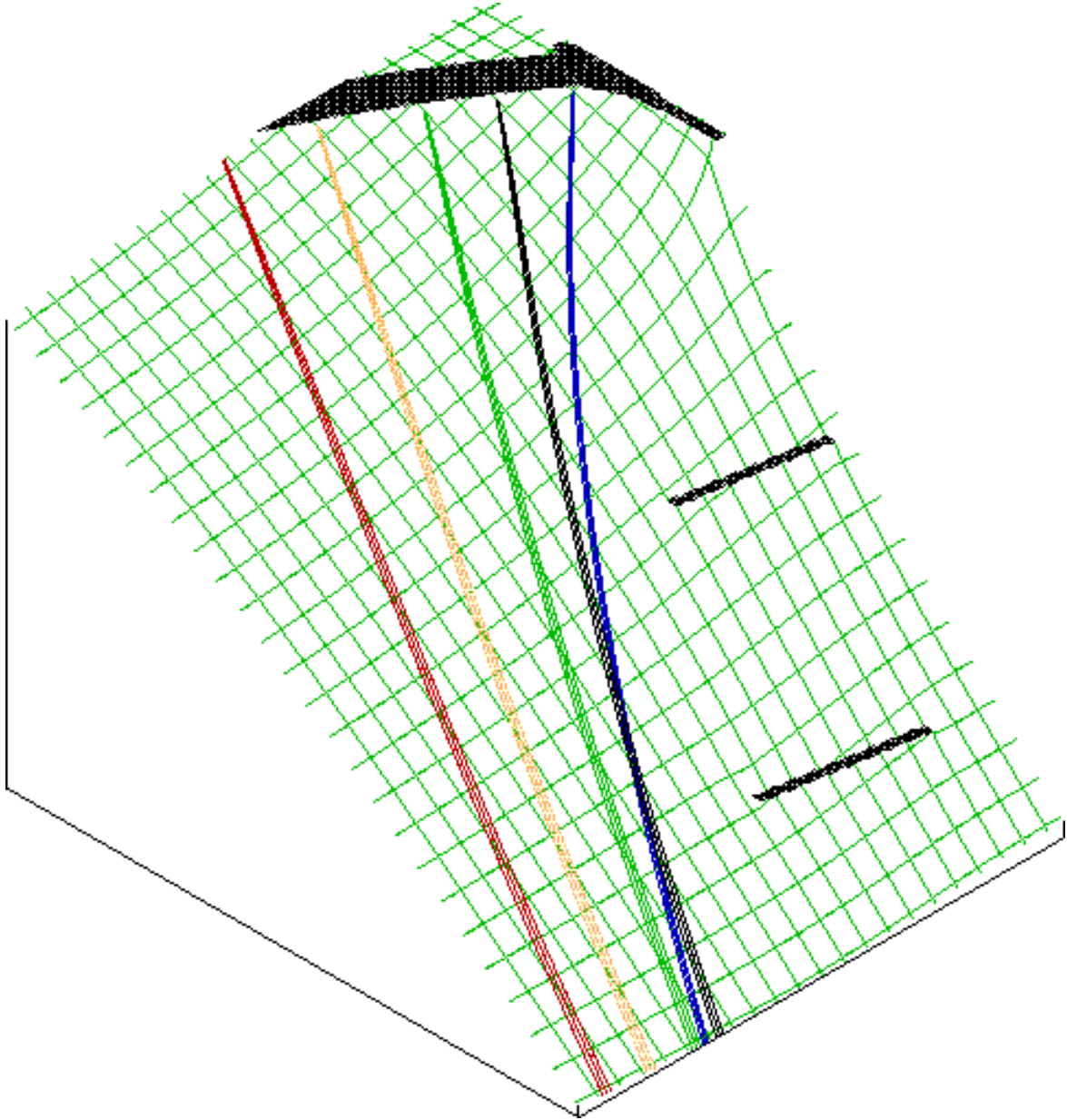,bbllx=50,bblly=150,bburx=792,bbury=480,width=22cm}
\caption{
\noindent
Distribution of the potential (or potential energy) inside the
tube. This way of presentation is similar to the old rubber-sheet
deformation method that has been widely used to visualize potential
distributions.
Strong field curvature is evident in the region close to the
periphery of the photocathode (right upper corner of the figure).
}
\label{FPP021}
\end{figure}

The reason for the failure 
becomes evident from Fig.~\ref{FPP02} and  Fig.~\ref{FPP021}.
There is a rapid change in the curvature of
the equipotential surfaces 
as the periphery of the tube is approached,
apparently due to
the shape of the conducting surface.
Photoelectrons emitted
from that region are directed too strongly 
towards
the center of HPD.

To fix this problem, one should reduce the curvature
of the potential distribution. Ideally one would prefer to
have equal curvature across the entire photocathode, and would be
therefore tempted to redesign the window supporting
structure.
However, for constructional reasons - precisely those related
to the achievement of the minimum constructional dead area -
this was not possible.

\begin{figure}[*]
\epsfig{file=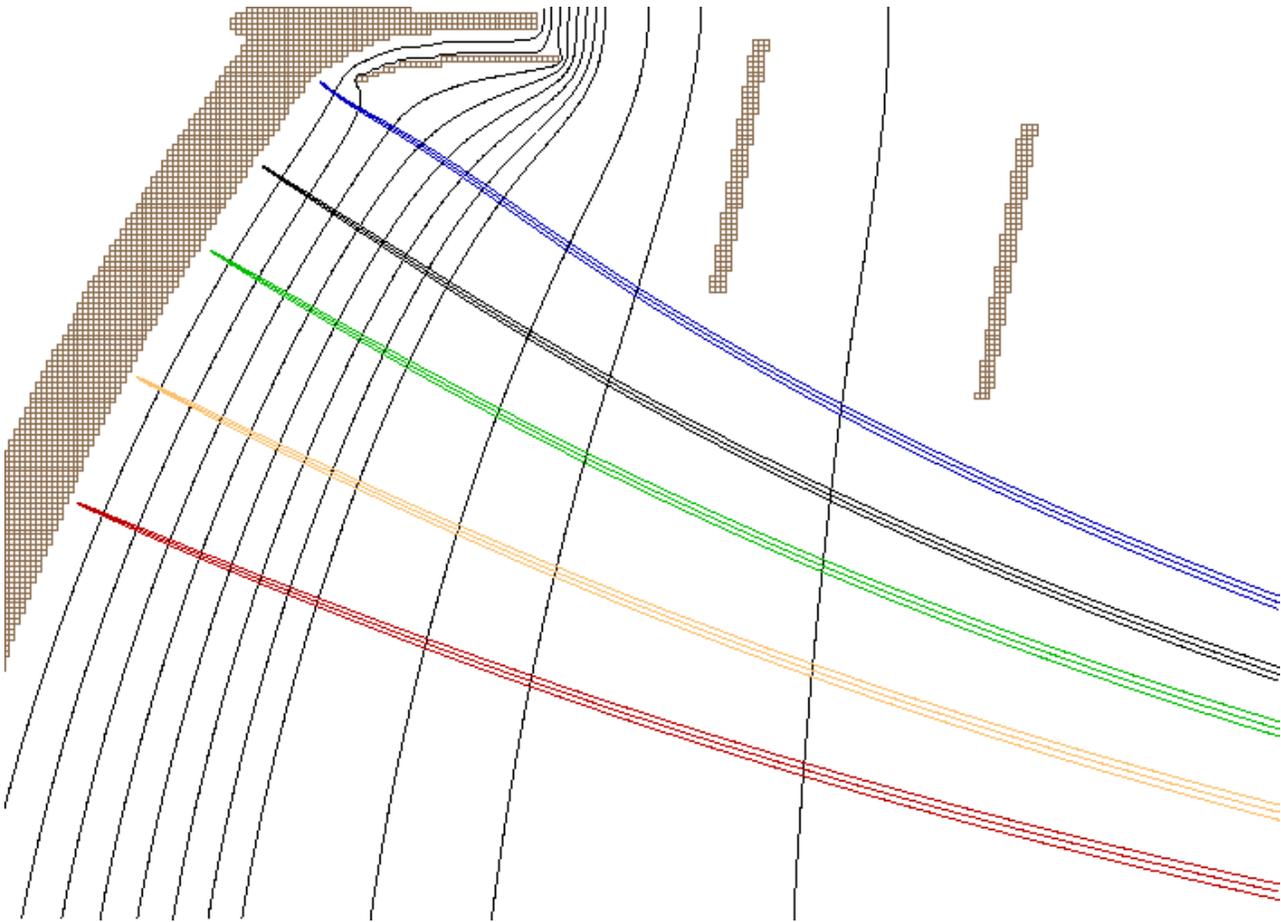,bbllx=50,bblly=200,bburx=792,bbury=480,width=24cm}
\caption{
\noindent
A cure for the strong curvature of the equipotential lines seen in 
Fig.~\ref{FPP01},~\ref{FPP02} and ~\ref{FPP021}
was found in the insertion of a new electrode - the ``bleeder electrode".
Surplus potential lines are conducted out from the HPD through
the slot created between the bleeder electrode and the body of the tube. 
}
\label{FPP03}
\end{figure}

\begin{figure}[*]
\epsfig{file=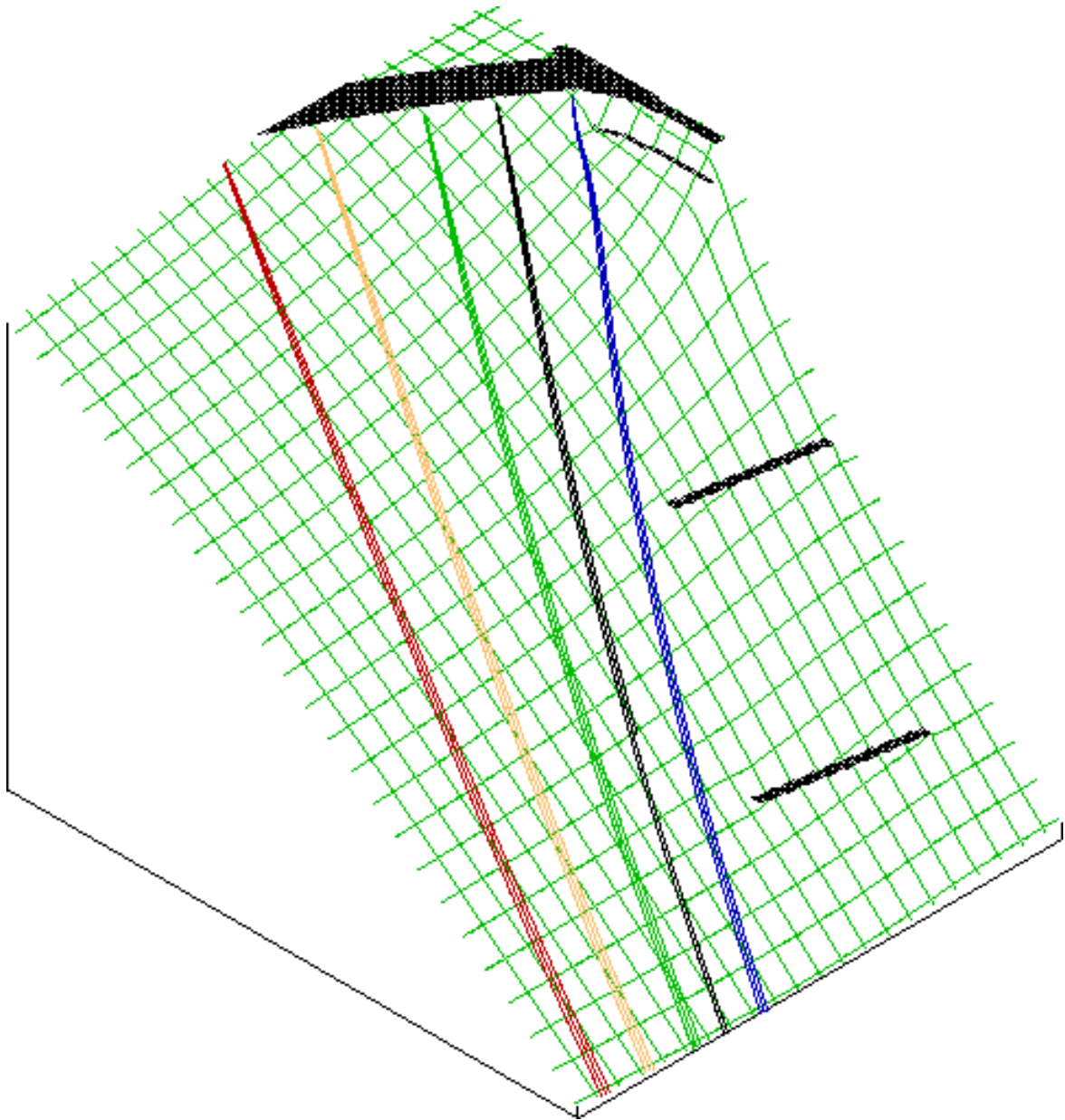,bbllx=50,bblly=150,bburx=792,bbury=480,width=22cm}
\caption{
\noindent
The bleeder electrode relaxes
the strong field curvature considered in Fig.~\ref{FPP021} 
and enables correct focusing.
}
\label{FPP031}
\end{figure}

\begin{figure}[*]
\epsfig{file=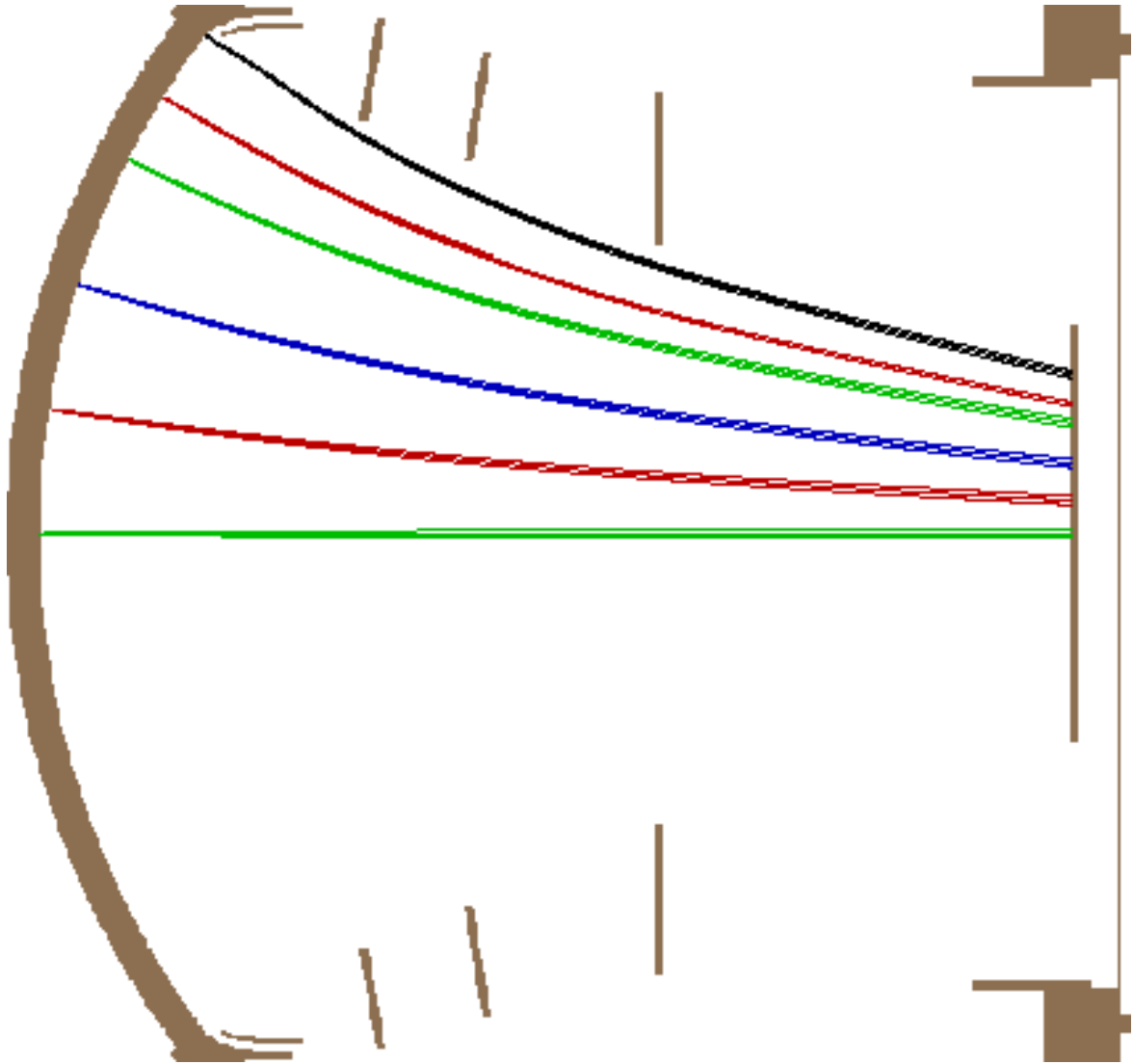,bbllx=50,bblly=150,bburx=792,bbury=620,width=22cm}
\caption{
\noindent
Thanks to the bleeder electrode,
the entire photocathode 
maps correctly onto the silicon
sensor. 
Electrodes are kept at the following potentials, from left to
right, respectively: -20 kV, -19.45 kV, -15 kV, -11 kV, -4.7 kV and 0 V.
}
\label{FPP04}
\end{figure}

We have  searched therefore for another solution
with the idea to reduce the field curvature
by ``conducting" some of the potential lines 
out from the 
tube, around the metallic window support.
The solution
was found in the creation of a slot which 
acts as a ``potential-conductor", see Figs.~\ref{FPP03},~\ref{FPP031} and~\ref{FPP04}. 
The slot was created by the insertion of a specially shaped
new electrode, the so called ``bleeder electrode"
(since it bleeds off the ``unwanted" field).
The unwanted potential lines are indeed conducted away
through the slot between the bleeder electrode 
and the body of the tube, and
the resulting field
in the problematic peripheral region has evidently lost its
strong curvature, as presented in
Figs.~\ref{FPP03},~\ref{FPP031} and~\ref{FPP04}.
The mapping quality is presented in Fig.~\ref{FPP011},
as a function of the potential on the bleeder electrode.
Since the acceptance is extended out to 60 mm away
from the tube axis, 63.5 mm being the outer tube radius,
about 89\% of the total photocathode surface now maps
correctly onto the silicon-pad 
sensor, and the functionally dead area got 
considerably reduced.

\begin{figure}[*]
\epsfig{file=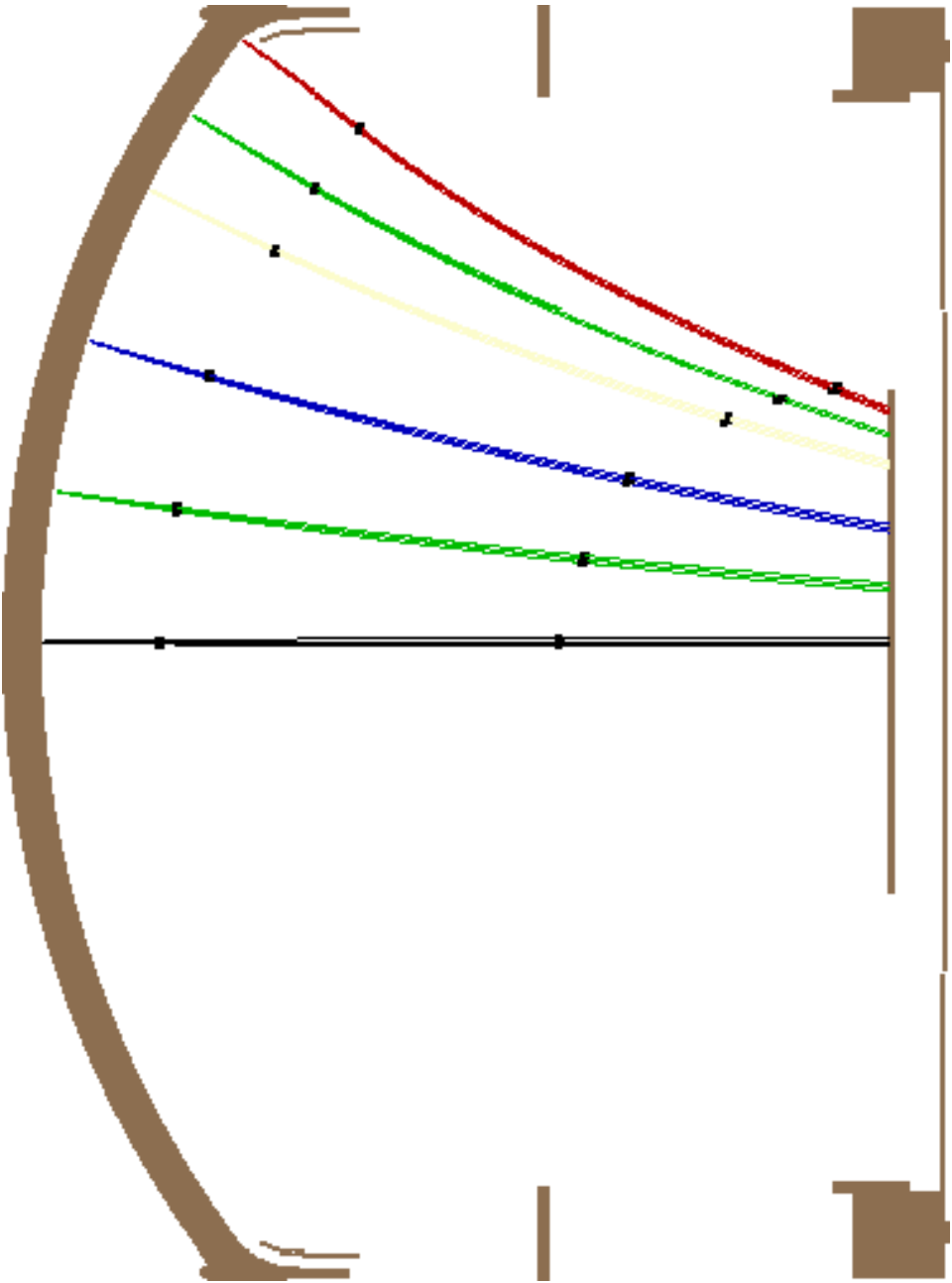,bbllx=50,bblly=50,bburx=792,bbury=620,width=18cm}
\caption{
\noindent
A considerably shorter version of the 5-inch HPD demonstrates 
a faster time response, and should be less
sensitive to residual magnetic fields.
Time marks of 1 ns are
drawn along the trajectories. 
Electrodes are kept at the following potentials, from left to
right, respectively: -20 kV, -19.6 kV, -13 kV and 0 V.
}
\label{FPP041}
\end{figure}

\begin{figure}[*]
\epsfig{file=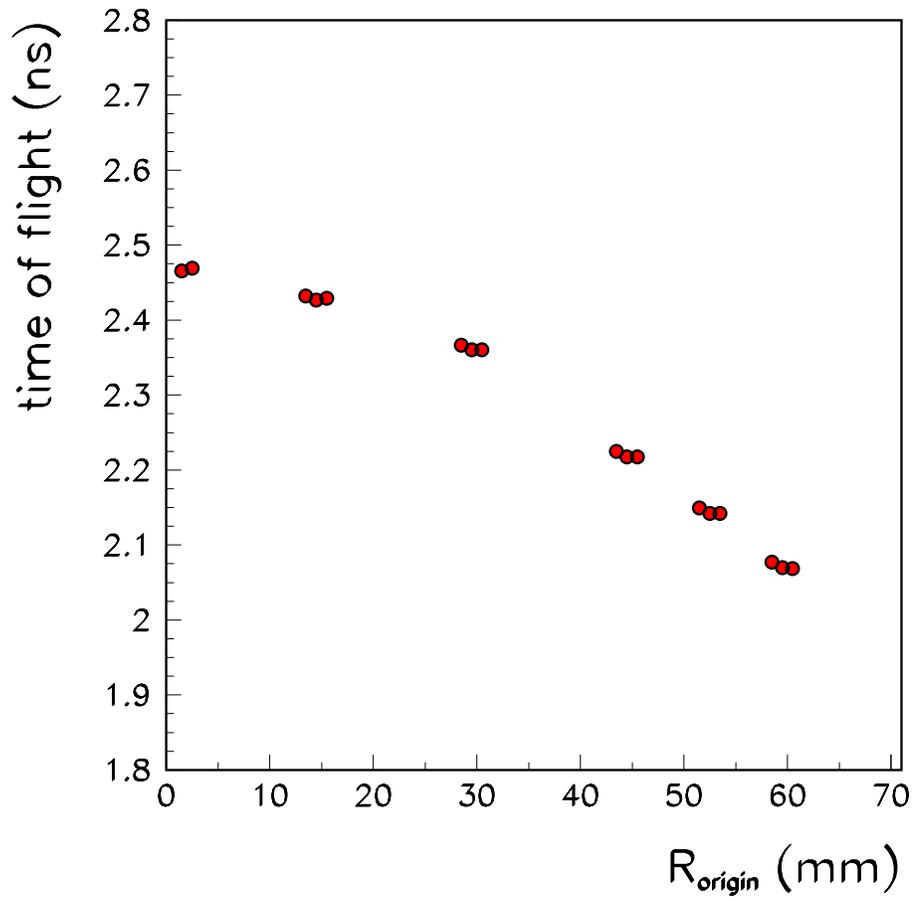,bbllx=50,bblly=50,bburx=792,bbury=620,width=18cm}
\caption{
\noindent
Time of flight in the short HPD from Fig.~\ref{FPP041}
as a function of R$_{origin}$.
}
\label{FPP0411}
\end{figure}

It is interesting to note that a considerably shorter HPD, but
of the same transverse size and entrance window,
may provide the same imaging accuracy, consider Fig.~\ref{FPP041}.
The  advantages of this short HPD are a very short time of flight,
as presented in Fig.~\ref{FPP0411},
and a weaker sensitivity to the residual magnetic
field. 
The time of flight is not only short, but the
spread corresponding to a given point of origin
on the photocathode is very narrow, as demonstrated
by triplets in Fig.~\ref{FPP0411}.
The weak sensitivity to the residual
magnetic field is due to the lower B~L 
and to the very rapid electron acceleration, since
the magnetic radius of curvature is proportional 
to the velocity.
A certain disadvantage may be 
the relative closeness of the 
electrodes with a large potential difference.

\begin{figure}[*]
\epsfig{file=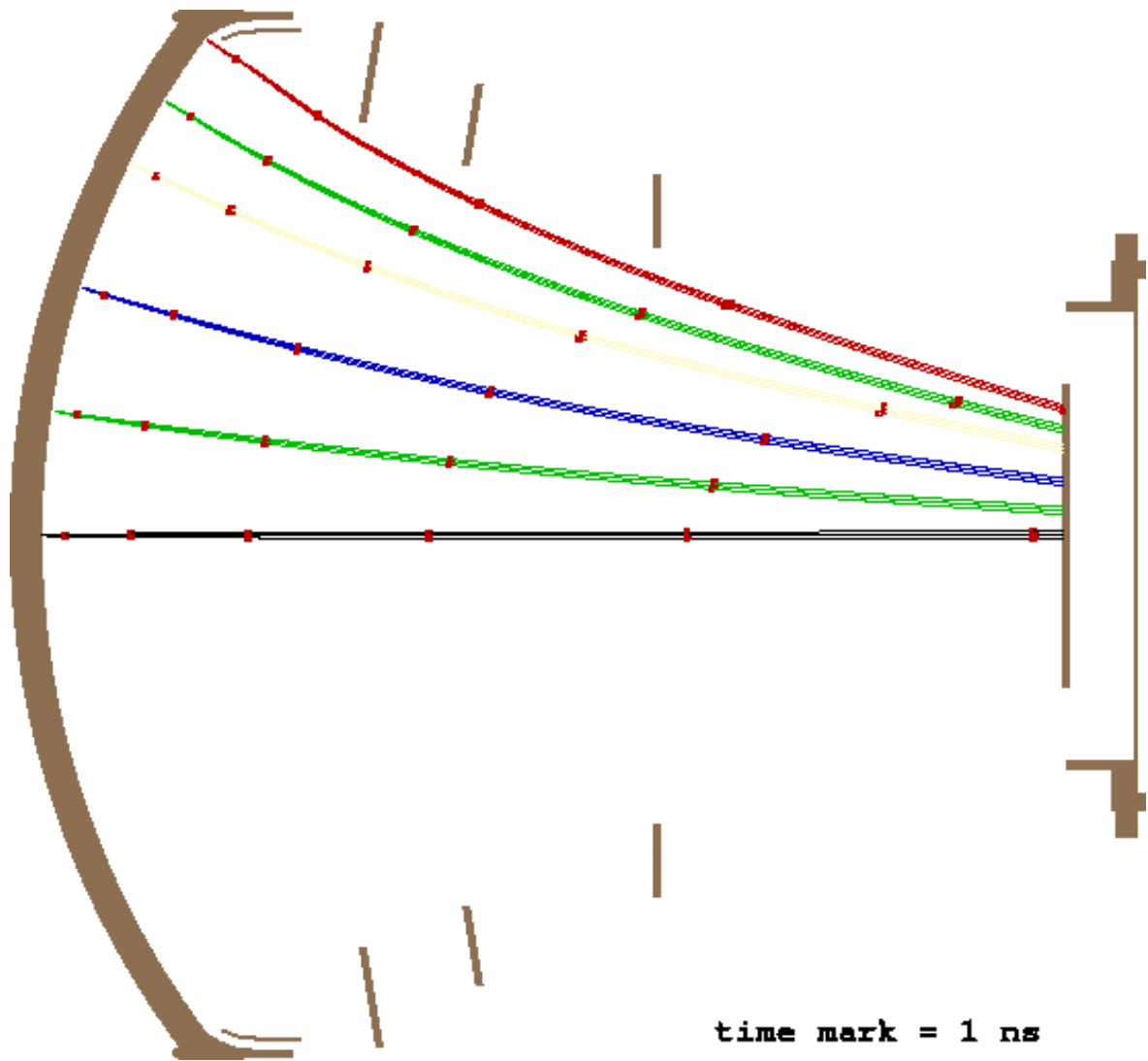,bbllx=50,bblly=120,bburx=792,bbury=620,width=22cm}
\caption{
\noindent
20-inch diameter HPD for the AQUA-RICH experiment. 
Electrodes are kept at the following potentials, from left to
right, respectively: -45 kV, -44.25 kV, -37.5 kV, -33 kV, -13.5 kV and 0 V.
Marks along the 
electron trajectories indicate time intervals of
1 ns.
}
\label{FPPAQUA}
\end{figure}

Excellent imaging
properties achieved by this method are equally welcome in
stand-alone devices. 
Therefore the same method has been applied in the design of
the 20-inch HPD tube (see Fig.~\ref{FPPAQUA}) for the 
AQUA-RICH neutrino experiment, proposed for
the CERN--Gran Sasso neutrino beam programme~\cite{aquarich}.
These HPDs are supposed to be
placed in a array with a large distance between units.
The AQUA-RICH HPDs should detect Cherenkov ring images 
from charged products of neutrino interactions
in a huge water tank. The silicon sensor will 
consist of 85 pixels.
A stronger demagnification (relative to the LHCb design)
was needed
because of the
large tube diameter, and the tube length was therefore
increased accordingly. The first focusing electrode that follows
the bleeder electrode plays in this design a very active role
in appropriate imaging, which was not the case in the 
LHCb tube design,
where essentially the same imaging performance
could be achieved without
any focusing electrode
except the bleeder (the other electrodes served
to assure a
stable potential distribution).

Timing properties, i.e. the distribution of arrival times of
photoelectrons to the semiconductor sensor, will be 
very important for the AQUA-RICH experiment since it is supposed to 
supply detailed information about the real-time Cherenkov
pattern development. 
Due to the strong potential gradient close to the photocathode, 
this device has a relatively good timing
characteristics. The time evolution of the electron
trajectories 
is indicated by 1 ns time marks along the 
trajectories
in Fig.~\ref{FPPAQUA}. Note however that further optimization
is possible. For example, a shorter tube should be considered,
like the short tube in Fig.~\ref{FPP041}.

To summarize, we have presented HPD designs essentially based on 
the application of the bleeder electrode.
By suitably changing the
field configuration in the critical region close to the
periphery of the photocathode, the bleeder electrode provides
correct imaging and considerably reduces the functional
dead area in large imaging HPDs.
Out of the three designs presented, the long tube for 
LHCb has been produced and recently successfully tested~\cite{HPD-test},
and the HPD tube for AQUA-RICH has been ordered for production
in a downscaled size 
(10-inch diameter)
for real physics tests.

The same method has been successfully applied in yet another
HPD design,
which incorporates also the
protection against the positive ion feedback.
That is the second topic of this publication
and the subject of the following section.

\section{Ion Feedback Protection in Large-area Imaging HPDs}

The presence of positive ions in the PMT and HPD vacuum
tubes presents one of the main drawbacks,
particularly in applications with high 
background or low signal photon rates, like in 
Atmospheric Cherenkov Telescopes (ACT) in the field
of gamma ray astronomy~\cite{NIM,Razmik,Eckart-NIM},
or in Cherenkov neutrino experiments.
Acceleration and subsequent
dumping of positive ions into a
photocathode leads both
to creation of noise
through electrons released,
and
to a rapid damage of the
photocathode
~\cite{NIM,Razmik,Eckart-NIM}.
In high-vacuum tubes the vast majority of
positive ions do not originate from residual gas, 
but
rather from the
the surface of the anode,
upon the impact of
accelerated photoelectrons.
Apart from Oxygen and Hydrogen atoms from adsorbed 
water, Cesium atoms are particularly
abundant because they usually
spread inside tubes during
and after the
manufacturing (activation) of photocathodes.

General solution to the problem of the
feedback of positive ions emerging from such processes
was found~\cite{NIM} in 
the 
insertion of a permanent electrostatic potential barrier
in front of the surface of the Photo Diode (PD) in HPDs,
or in front of the first dynode in PMTs.
This method was originally proposed ~\cite{NIM} to modify
the small (2-cm-diameter) single-pixel Intevac
HPD~\cite{Eckart-NIM}. 
In this paper we demonstrate that the same idea may
be successfully applied to a large area imaging HPD.

The potential barrier is created by the so called
``barrier electrode"
placed in front of the silicon sensor 
and kept at a {\bf positive} potential with respect
to the anode. 
The potential needed 
to establish a sufficiently high potential barrier
strongly depends on the size of the opening in the
barrier electrode, which is in turn determined 
by the width of the beam of photoelectrons, in order
to allow a free passage of the photoelectrons.
The wider the opening, the higher the positive potential
on the electrode. 

The HPD tube designed for LHCb, presented
in the previous section, Fig.~\ref{FPP04}, is characterized by 
proximity focusing electron optics  
(i.e. the diameter of the electron pattern shrinks
progressively down to the silicon  sensor size). 
Due to the very wide electron pattern
the barrier electrode should have a wide opening,
and that would require a very
high positive potential in order to establish the barrier effect. 
The proximity focusing
design is therefore very unsuitable for the 
implementation of the positive ion feedback 
protection. 

\begin{figure}[*]
\epsfig{file=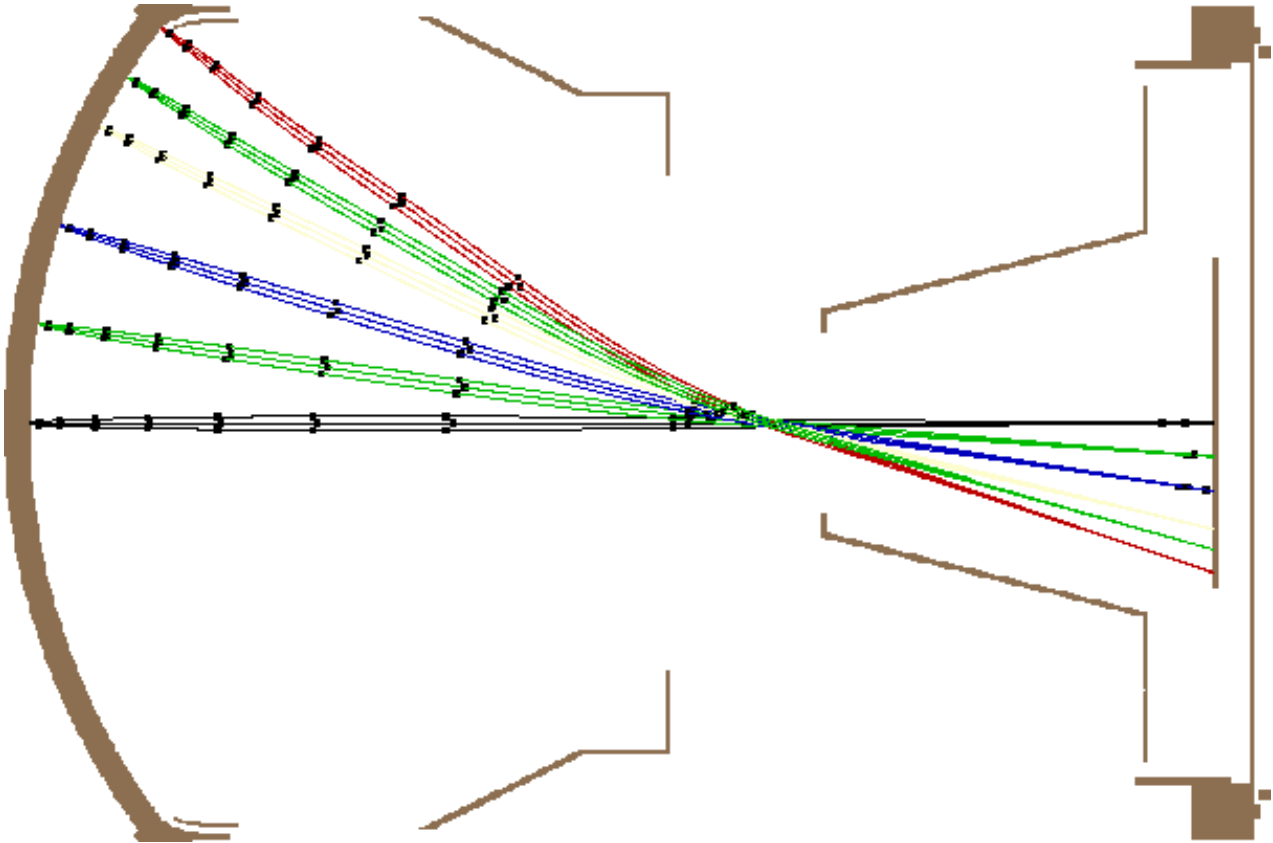,bbllx=50,bblly=150,bburx=792,bbury=620,width=24cm}
\caption{
\noindent
Cross-focusing 5-inch diameter HPD, with superior imaging characteristics
and positive ion feedback protection.
Electrodes are kept at the following potentials, from left to
right, respectively: -20 kV, -19.97 kV, -19.4 kV, {\bf +100 V}, and 0 V.
Marks along the
electron trajectories indicate time intervals of
1 ns.
}
\label{FPP1}
\end{figure}

However, a modified HPD design 
(see Fig.~\ref{FPP1})
with the so called ``cross-focusing" electron optics
provides the means to apply our positive ion 
protection method. The point is that electron
trajectories originating at any point on the photocathode
cross in a single place about the middle of the 
tube, and it is possible to insert a barrier electrode
with a very narrow opening, placed just at the crossing
point of the electron trajectories. There are many reasons
why this electrode should have a conical shape, as in Fig.~\ref{FPP1},
like e.g. the need to separate the connection
to the power supply outside the tube as far as possible from the 
throughput of the 
neighbour electrode with opposite polarity.

\subsection{Cross Focusing Electron Optics}

\begin{figure}[*]
\epsfig{file=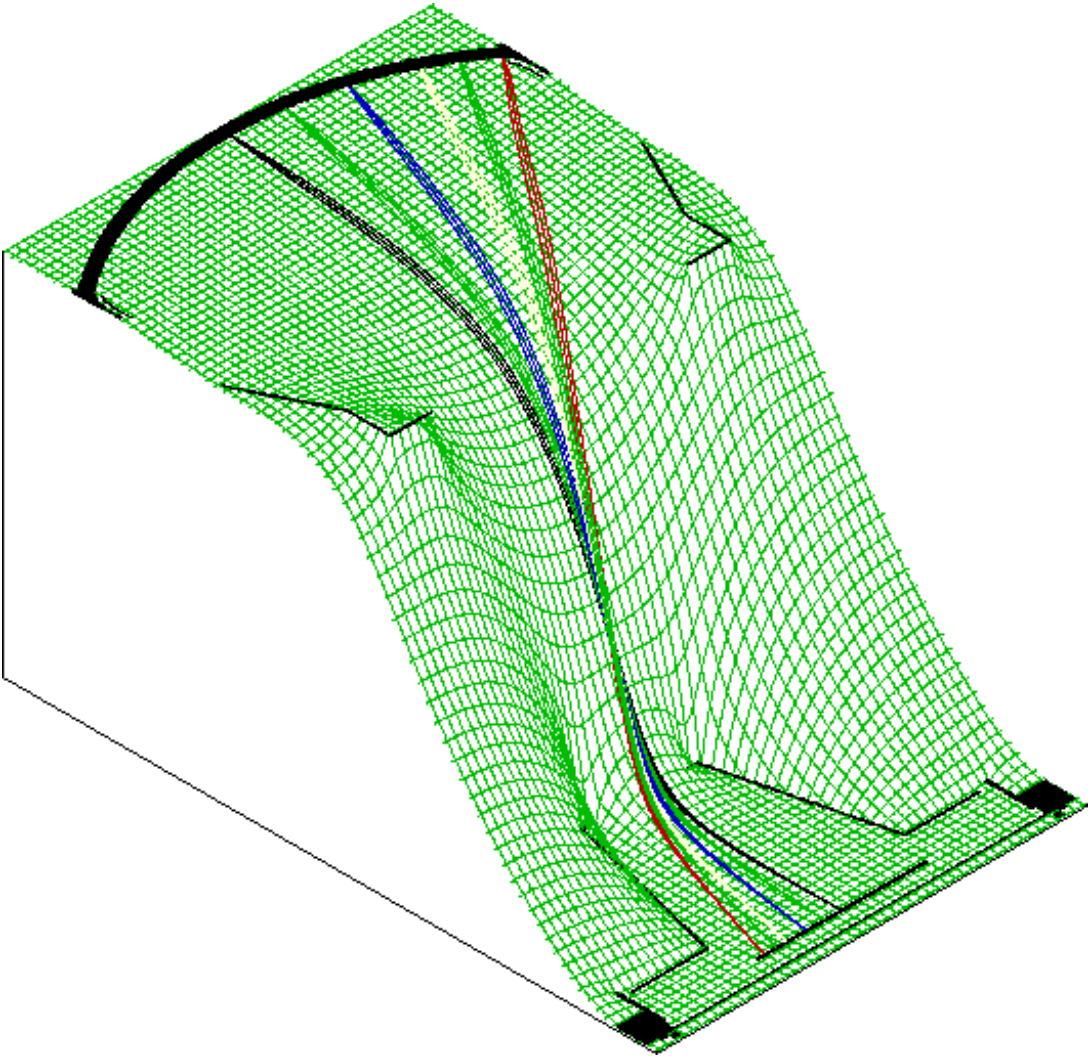,bbllx=50,bblly=200,bburx=705,bbury=520,width=18cm}
\caption{
\noindent
A view on the potential distribution in HPD from Fig.~\ref{FPP1}.
The interior of the conical electrode is essentially
field-free and the electron trajectories are ``frozen" 
at the entrance.
}
\label{FPP102}
\end{figure}

\begin{figure}[*]
\epsfig{file=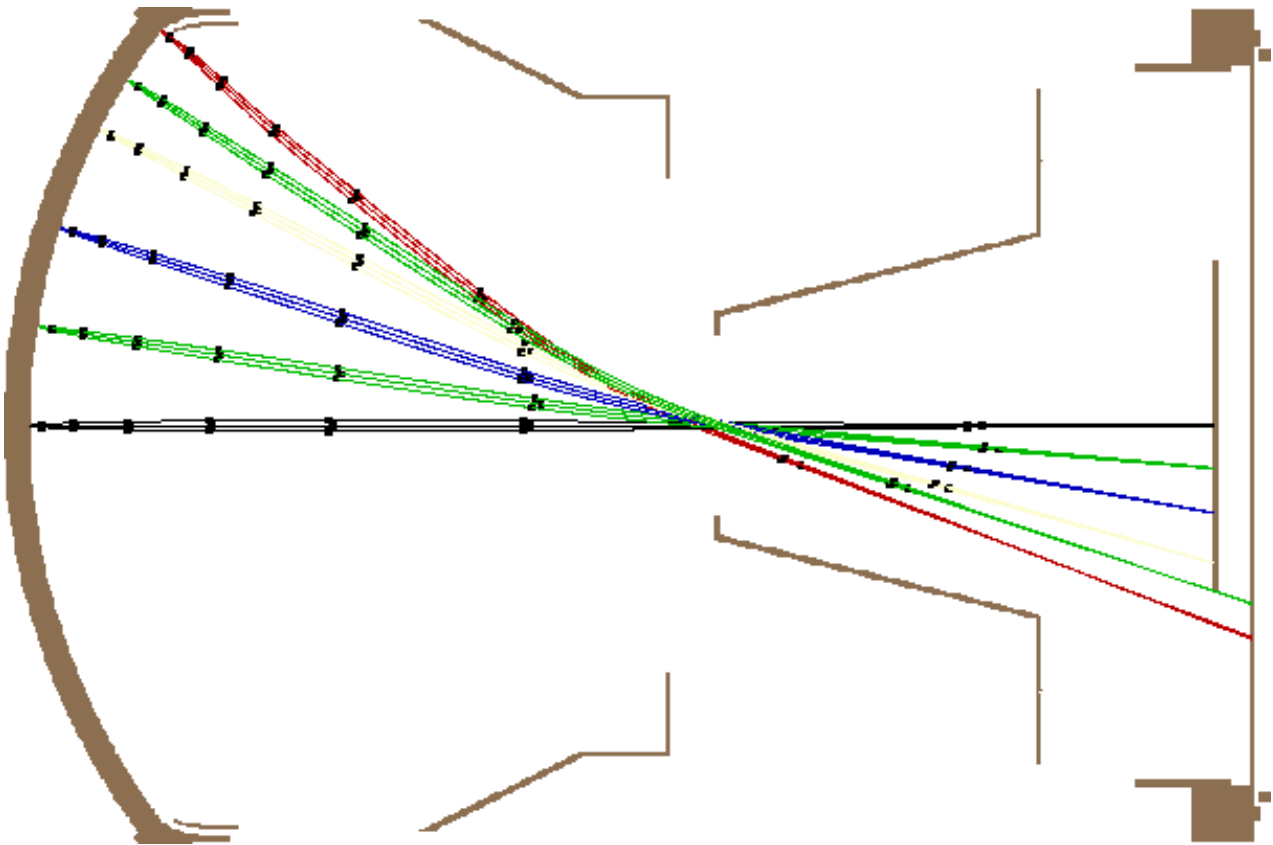,bbllx=50,bblly=150,bburx=792,bbury=620,width=24cm}
\caption{
\noindent
The ``zooming" functionality of the 
conical electrode in the electron lens --
the size of the projected image has increased
after 
the 
conical electrode was moved closer to 
the photocathode.
}
\label{FPP101}
\end{figure}

Before we turn to the discussion of the functionality
of the conical electrode for ion protection, let us
first review its functionality in electron optics of
a cross focused device.
Roughly speaking,
the conical electrode 
acts as a ``zoom" in the electron lens. 
The space enclosed by
the conical electrode is practically field-free, see Fig.~\ref{FPP102},
and therefore from its entrance onwards electrons keep
their velocities and directions until they
hit the anode surface. 
The
zooming functionality of the conical electrode
is related precisely to this feature -
the electron velocity vectors are ``frozen"
after they enter the cone, consider Figs.~\ref{FPP102}, and placing
the cone entrance earlier or later along the 
electron trajectories 
freezes the electron momenta with more or less divergence,
respectively. This feature is demonstrated in 
Fig.~\ref{FPP101} -- the conical electrode was moved
closer to the photocathode, and consequently the image got 
magnified.

To achieve a good ``image sharpness", on the other hand,
one should optimize
the potential on the electrode placed between the 
conical electrode and the photocathode. 

In the development of the cross focused tube, presented in Fig.~\ref{FPP1},
we started 
from the proximity focused HPD design, presented in Fig.~\ref{FPP04},
and introduced
a number of significant modifications.
We kept the same diameter (5-inch) and
shape of the entrance window in order to be able to use
the same parts in tests. However, the tube
got considerably elongated to provide the means for
the cross-focusing electron optics. The conical electrode and the
electrode closer to the photocathode were introduced.

\begin{figure}[*]
\epsfig{file=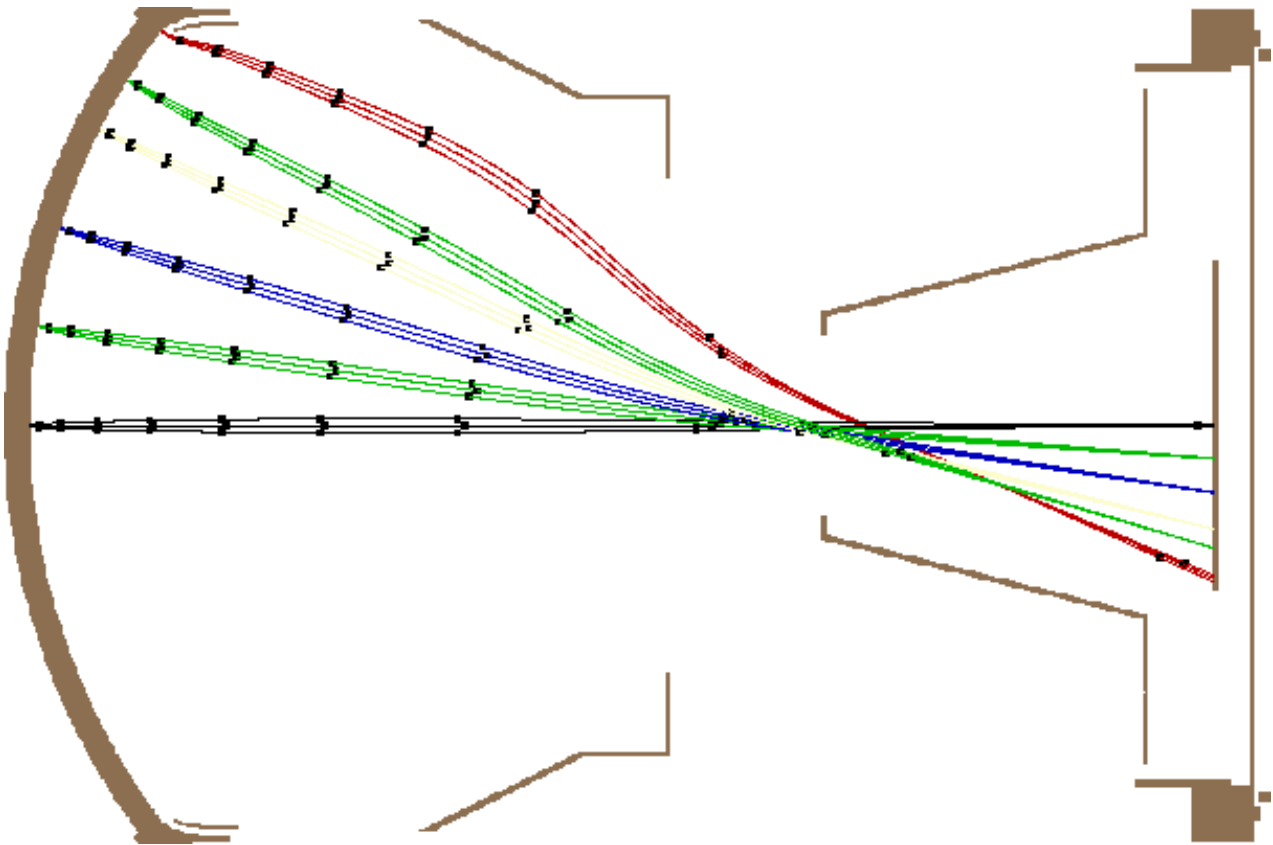,bbllx=50,bblly=150,bburx=792,bbury=620,width=24cm}
\caption{
\noindent
HPD from Fig.~\ref{FPP1} with modified
potential U(Bleeder Electrode) = -19.9 kV. 
Trajectories
originating close to the periphery of the photocathode 
demonstrate odd behavior.
Marks along the
electron trajectories indicate time intervals of
1 ns.
}
\label{FPP12}
\end{figure}

\begin{figure}[*]
\epsfig{file=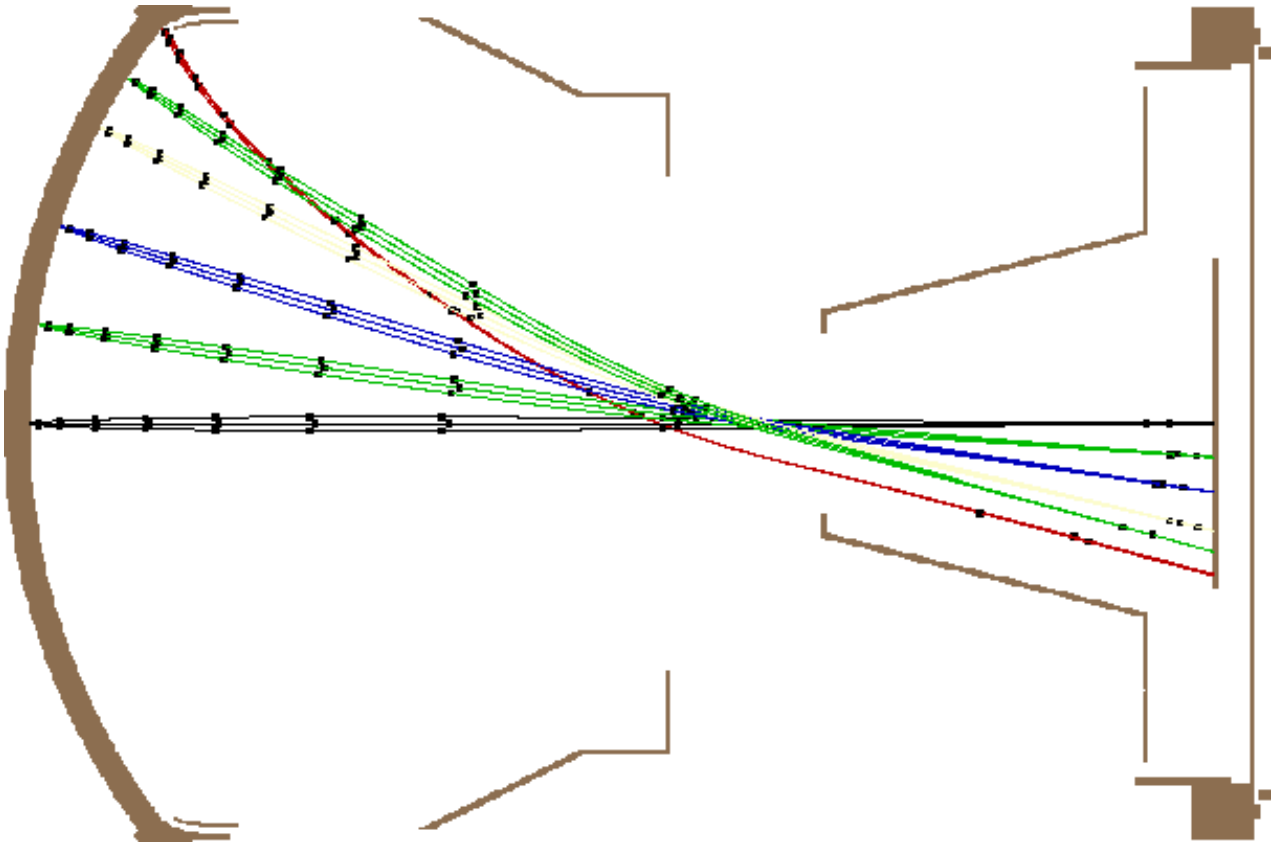,bbllx=50,bblly=150,bburx=792,bbury=620,width=24cm}
\caption{
\noindent
HPD from Fig.~\ref{FPP1}, now with
potential U(Bleeder Electrode) = -20.0 kV.
Trajectories
originating close to the periphery of the photocathode 
again show odd behavior, but in the opposite direction 
than in Fig.~\ref{FPP12}.
Marks along the
electron trajectories indicate time intervals of
1 ns.
}
\label{FPP13}
\end{figure}

\begin{figure}[*]
\epsfig{file=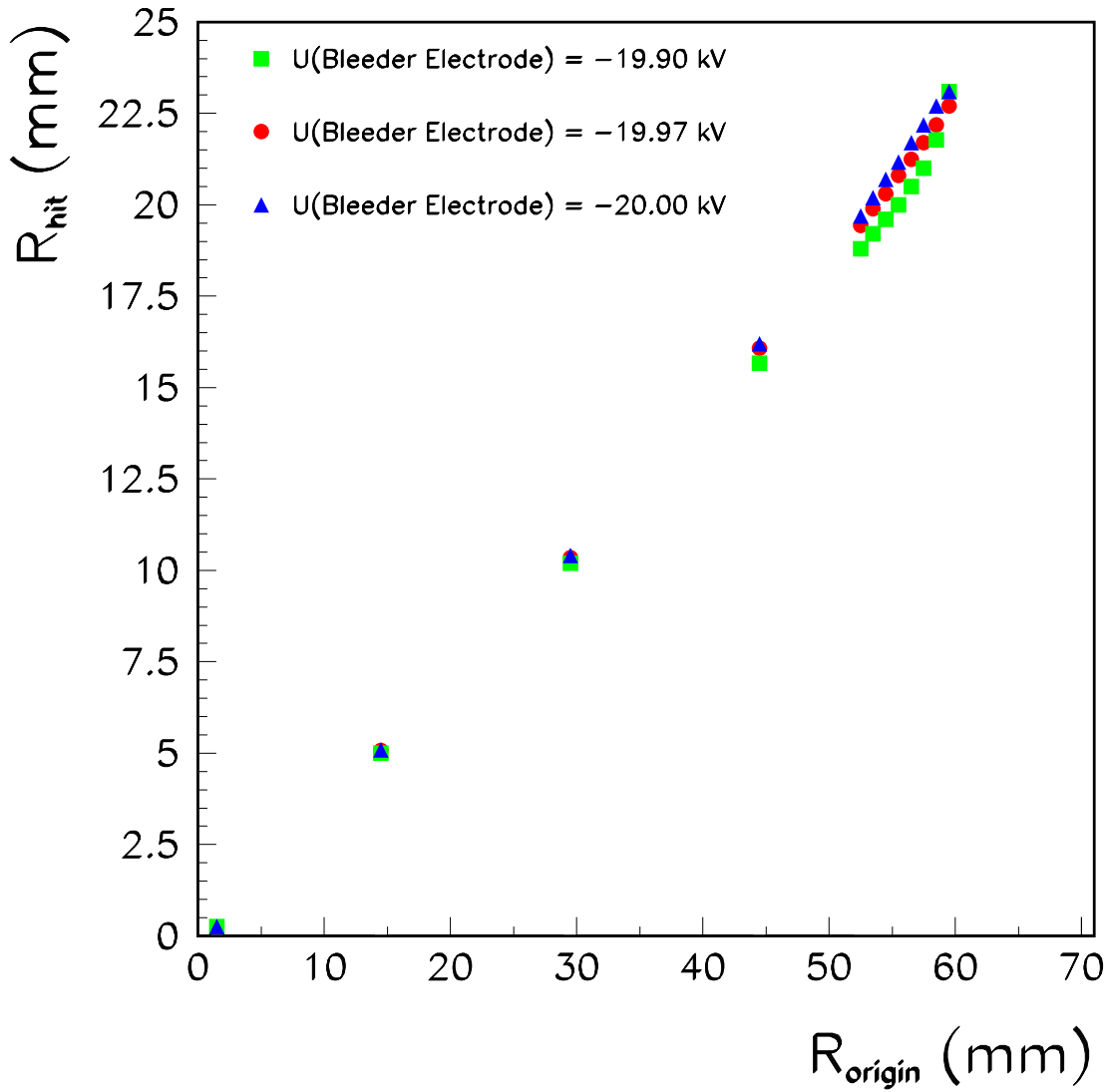,bbllx=50,bblly=200,bburx=792,bbury=480,width=22cm}
\caption{
\noindent
Image mapping from the photocathode to the 
silicon pad sensor. 
Electrons originate at radius R$_{origin}$ away
from the HPD axis
and hit the silicon sensor at R$_{hit}$. 
The bleeder electrode has a weaker
influence on mapping than in the proximity focusing HPD (Fig.~\ref{FPP011}),
but it strongly influences trajectories in their early 
stage, compare Figs.~\ref{FPP1},~\ref{FPP12}, and~\ref{FPP13}. 
}
\label{FPP11}
\end{figure}

\begin{figure}[*]
\epsfig{file=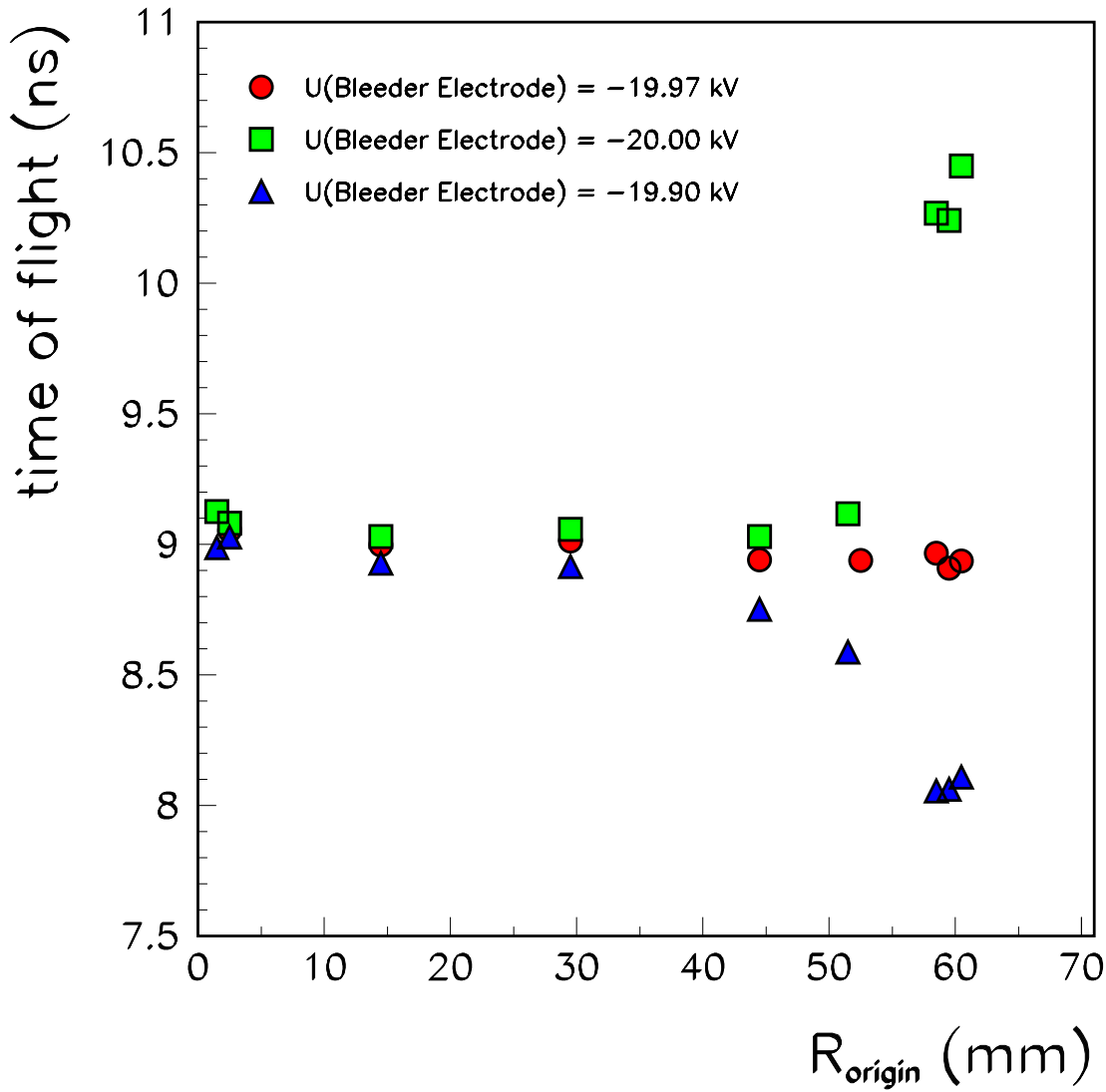,bbllx=50,bblly=200,bburx=792,bbury=480,width=22cm}
\caption{
\noindent
Time of flight in the cross focused HPD 
as a function of R$_{origin}$,
for different potential settings on the bleeder
electrode. 
The bleeder electrode has a weaker
influence on mapping than in the proximity 
focusing HPD (Fig.~\ref{FPP011}),
but it strongly influences trajectories in their early
stage, 
and therefore also the timing properties;
compare Figs.~\ref{FPP1},~\ref{FPP12}, and~\ref{FPP13}.
}
\label{FPP111}
\end{figure}

The bleeder electrode stayed in place with essentially 
the same functionality. The effects of potential variations
on the bleeder electrode are presented 
in Figs.~\ref{FPP12},~\ref{FPP13},
~\ref{FPP11} and ~\ref{FPP111}. It is evident that the bleeder electrode plays 
again an important role in the shaping of the trajectories,
although the mapping is more robust than in the proximity
focusing HPD. A very coherent time evolution for
all trajectories may be achieved by careful tuning, as demonstrated
in Figs.~\ref{FPP1} and~\ref{FPP111}. Compared to the 
short proximity focused HPD (Figs.~\ref{FPP041} and ~\ref{FPP0411}),
the optimized cross focused device offers a much more 
coherent time evolution over the entire area,
while the total time of flight is around four times
longer, and the time spread for a given point of origin
$R_{origin}$ is wider. 

\subsection{Ion Feedback Protection}

\begin{figure}[*]
\epsfig{file=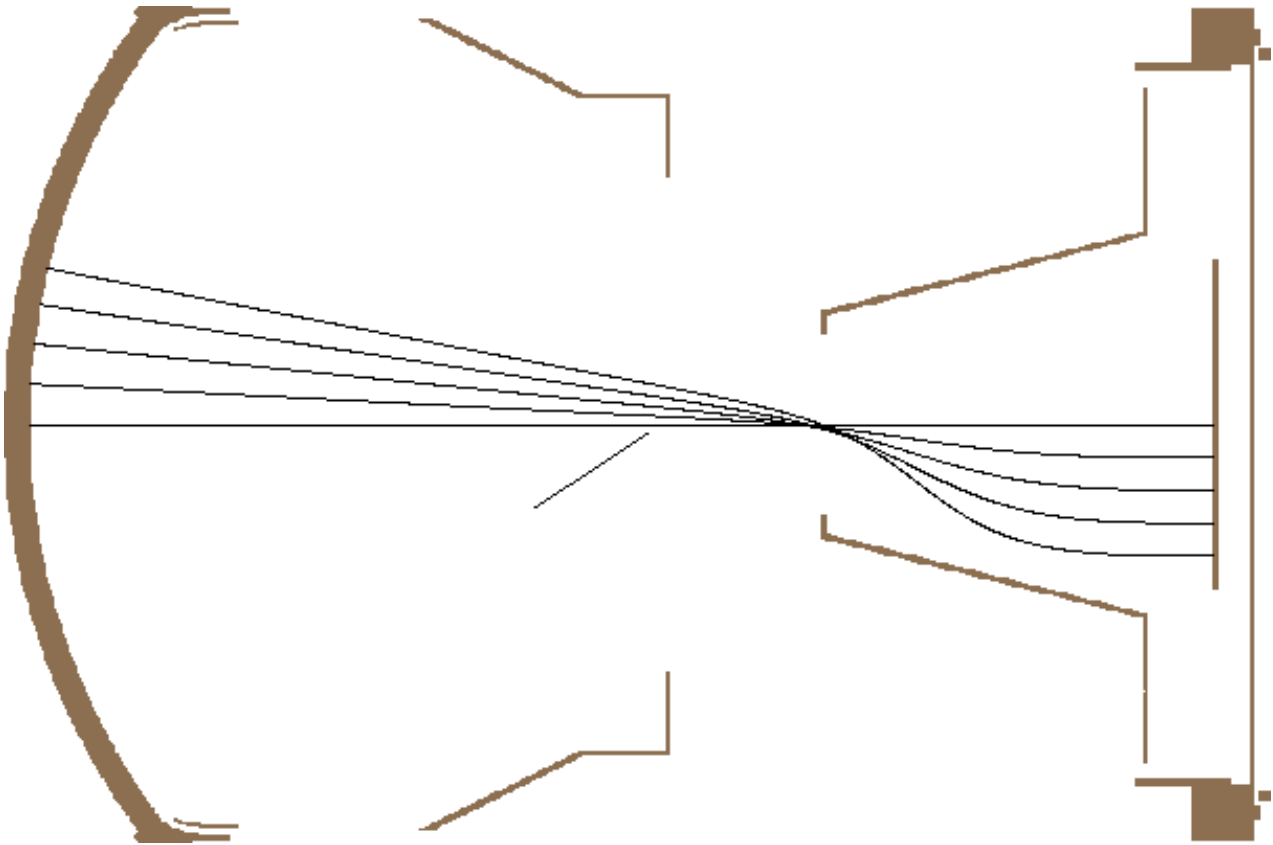,bbllx=50,bblly=150,bburx=792,bbury=620,width=24cm}
\caption{
\noindent
Ion feedback protection set ``off" -- conical
electrode at the anode potential (U=0). 
Ions
emerge from the surface of the anode (right) and accelerate
towards the photocathode (left), causing damage
and operational noise.
}
\label{FPP2}
\end{figure}

To demonstrate the positive ion feedback,
in Fig.~\ref{FPP2}
the conical electrode 
is kept at the anode potential
i.e. the ion protection was set ``off".
Trajectories of singly charged positive
ions are shown,
emerging with energies of 44 eV
at normal incidence
from the anode surface on the right side of the figure,
and moving towards the 
photocathode on the left side.

Note that the angular and
energetic distributions of positive ions
are, to our best knowledge, unknown. We have worked out a
scheme how to perform a measurement of those quantities,
using actually a tube with a barrier electrode,
but
since the results are not yet available,
we are currently using a rough
estimate that the ions could reach an energy of around 40 eV.
Once the actual energy will be measured,
it will be straightforward to repeat the
simulations and find the optimal potential settings.

\begin{figure}[*]
\epsfig{file=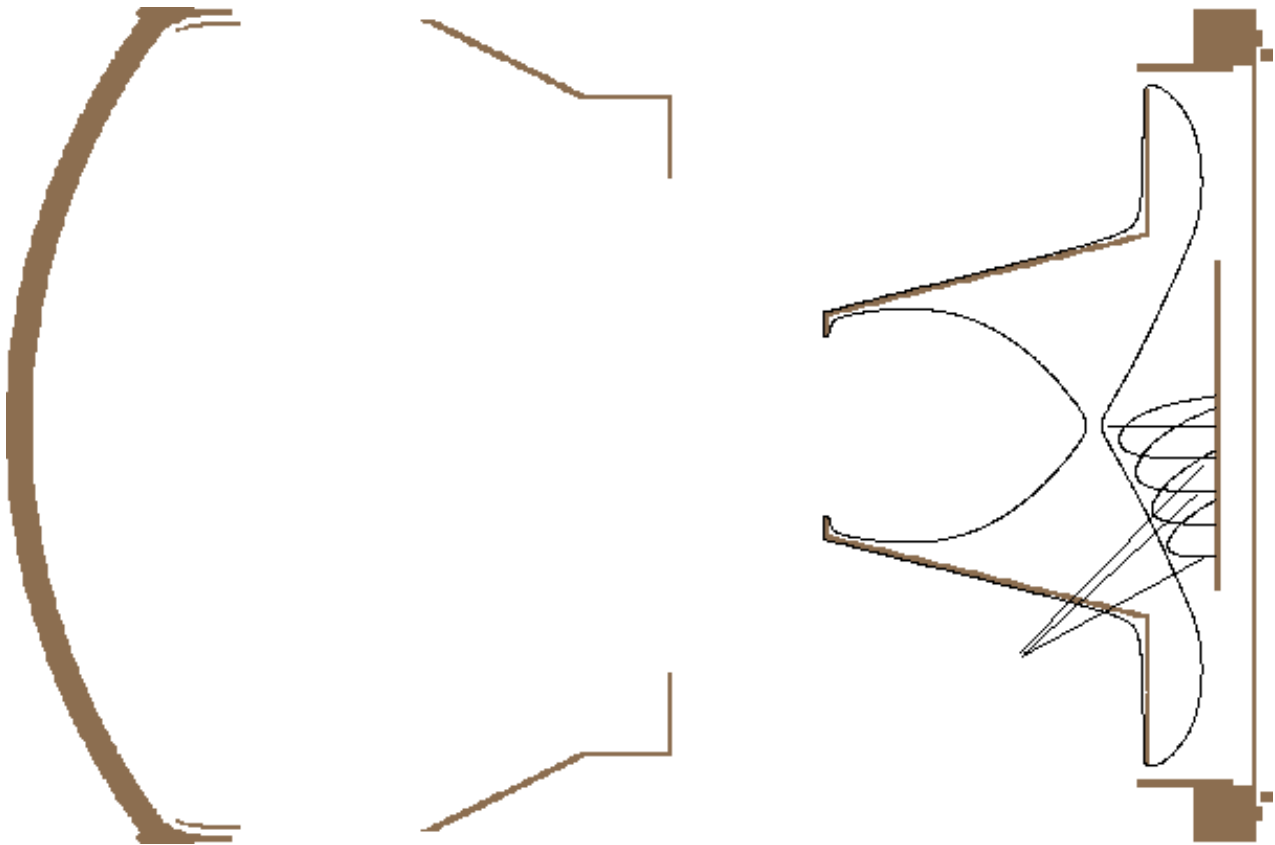,bbllx=50,bblly=150,bburx=792,bbury=620,width=24cm}
\caption{
\noindent
Ion feedback protection ``on" --
conical barrier  electrode at
+100 V. Between the barrier  electrode
and the anode a potential barrier is established 
to
repel back positive ions emerging from the
anode surface.
}
\label{FPP3}
\end{figure}

\begin{figure}[*]
 \epsfig{file=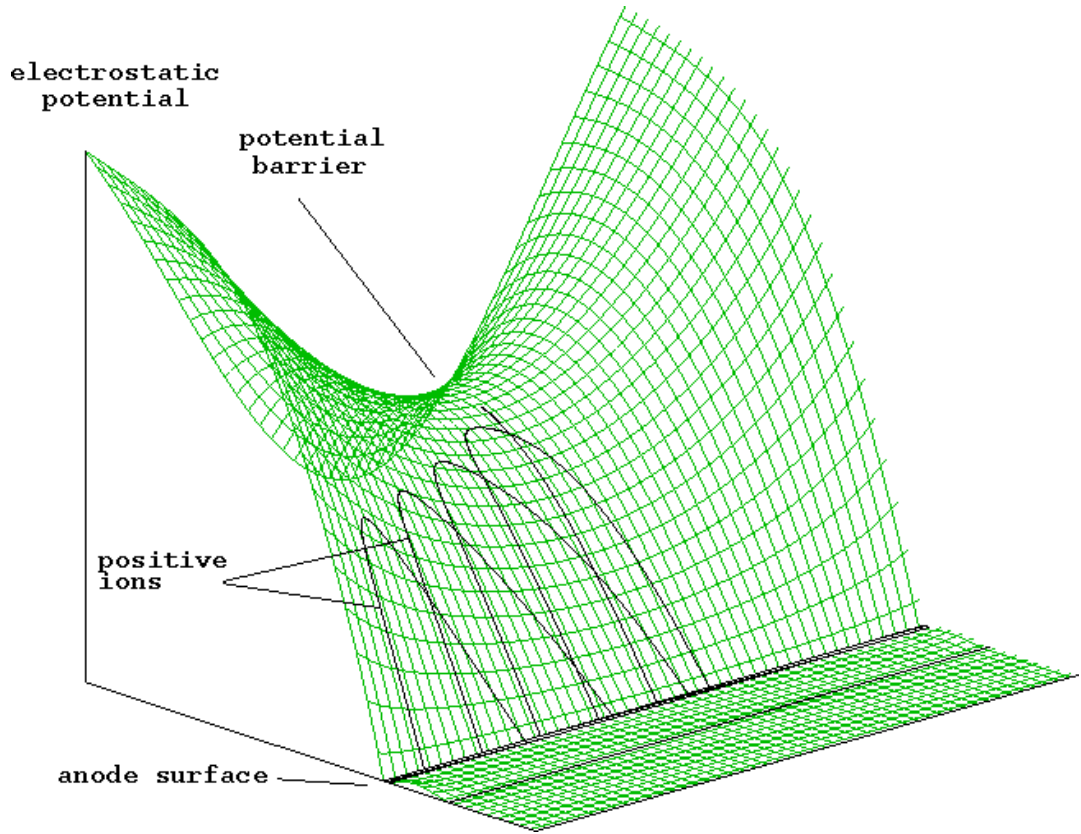,bbllx=50,bblly=100,bburx=792,bbury=500,width=20cm}
\caption{
\noindent
Ion feedback protection.
Potential distribution in front of the
anode plane of HPD from
Fig.~\ref{FPP3}.
Positive ions of energy E$_{ion}$=44 eV
and emission angle normal to the anode surface
start ``climbing" the potential barrier (E$_{b}$=45 eV)
but become repelled back and therefore never reach the 
photocathode.
}
\label{FPP4}
\end{figure}

The conical electrode starts to play the role
of the
``barrier electrode"~\cite{NIM}
when a sufficiently high positive potential is applied.
As demonstrated in Fig.~\ref{FPP3},
monotonous 
decrease of the potential for positive particles towards the 
photocathode
breaks down, and
a potential barrier (for positive particles) is created
in front of the anode. This barrier
prevents positive 
ions from penetrating further towards the photocathode
and solves the ion feedback problem.
For more clarity the potential 
distribution in front of the anode plane 
is shown in a magnified view in Fig.~\ref{FPP4}.
Trajectories of singly charged positive
ions are simulated with identical initial conditions
like before. Ions of energy lower than the 
barrier always get repelled back from the barrier and
never can reach the photocathode.

As argued already in~\cite{NIM}, the precision
of the potential on the barrier electrode which is
required
for stable electron focusing is
not a critical issue
-- variations of even 10\% on the potential will leave
the electron focusing essentially
unchanged.
An independent, very common
voltage supply may be therefore used
to bias the positive conical barrier electrode.

Apart from providing for the positive ion protection,
this tube is characterized by a much narrower spread
of photoelectrons on the silicon-pad sensor 
and therefore with a superior 
imaging performance.
There are also other 
benefits from the conical barrier electrode:
(i) it will capture a large fraction
of electrons back-scattered from the semiconductor
sensor, and (ii) to some extent it will
protect the anode area
from pollution during (and after) the manufacturing
of the tube, in particular in case of
in-situ photocathode activation. 

 Our cross focusing HPD should be tested
soon -- it was designed in a way to be rather easily assembled from
the elements
of
the proximity focusing HPD (Fig.~\ref{FPP04})
(entrance window with photocathode, cylindrical body, base plate and 
silicon sensor),
which have already been
produced.

The presented cross focused HPD, with all its virtues,
seems to be a very good choice for Atmospheric Cherenkov
Telescopes, Cherenkov neutrino detectors and in general
in applications which reqire good imaging, low noise
and coherent timing.

\section{Conclusions and Outlook}

Imaging
Hybrid Photon Detectors (HPD) have been developed
for integration in large area imaging Cherenkov
detectors to be used in high energy physics and astrophysics.
The presented designs have been developed
in particular for experiments MAGIC, LHCb and AQUA-RICH.
Apart from a very good imaging performance,
the presented HPDs
comprise
protection against positive ion feedback and(or)
minimal
dead area.
A detailed discussion of the underlying innovations
was presented.
We have introduced a new electrode (``bleeder") which
modifies the field configuration in the critical region close to the
periphery of the photocathode, thus
providing
correct image mapping and minimization of the 
dead area in large imaging HPDs. Three proximity
focused
and one cross focused
HPD design
were presented that are based on the
application of the bleeder electrode.
The latter design also incorporates the
protection against the positive ion feedback,
a feature very important for Atmospheric Cherenkov Telescopes,
but also for Cherenkov neutrino detectors.

So far, the presented HPD designs have already evolved beyond the 
design stage:\\
\noindent
(i) Proximity focusing HPDs for
LHCb (Fig.~\ref{FPP04}) were produced and
tested recently~\cite{HPD-test}. The results fully
confirm the presented design.\\
\noindent
(ii) Cross focusing HPDs (Fig.~\ref{FPP1}) should be assambled (essentially from
parts of proximity HPDs) and tested soon.\\
\noindent
(iii) Half-sized 
(10-inch diameter) 
HPDs for AQUA-RICH (Fig.~\ref{FPPAQUA}) 
are in production
for real physics tests in a water container exposed to 
a muon or hadron beam at CERN~\cite{MINI-AQUA}.

\section*{Acknowledgments}
Very stimulating discussions with 
Enrico Chesi, 
Eckart Lorenz,
Dario Hrupec,
Eugenio Nappi,
Guy Pai\' c and
Jacques Seguinot are gratefully acknowledged.
I am particularly grateful to Tom Ypsilantis
who also strongly motivated this publication.

\clearpage

\clearpage

\clearpage

\clearpage

\clearpage

\clearpage

\clearpage

\clearpage

\clearpage
\clearpage

\clearpage

\clearpage


\clearpage

\clearpage

\clearpage
\clearpage

\clearpage
\clearpage

\clearpage
\clearpage

\end{document}